\RequirePackage{ifpdf}

\documentclass{PoS}

 \usepackage{graphicx}
 \usepackage{epstopdf}
 \usepackage[authoryear,square]{natbib}
 \bibpunct{(}{)}{;}{a}{}{,}

\title{Synergy of CO/[CII]/Ly$\alpha$ Line Intensity Mapping with
  the SKA}

\ShortTitle{CO/[CII]/Ly$\alpha$ synergy}

\author{\speaker{Tzu-Ching Chang}$^1$,
  Yan Gong$^2$, Mario Santos$^{3,4,5}$, Marta Silva$^4$, James
  Aguirre$^6$, Olivier Dor\'e$^{7,8}$,
  Jonathan Pritchard$^9$, on behalf of the EoR/CD-SWG\\
        $^1$Academia Sinica Institute of Astronomy and Astrophysics,
        P.O. Box 23-141, Taipei, 10617 Taiwan \\
        $^2$Department of Physics \& Astronomy, University of
        California, Irvine, CA 92697, USA \\
        $^3$Department of Physics, University of Western Cape, Cape
        Town 7535, South Africa \\
        $^4$CENTRA, Instituto Superior T ́ecnico, Technical University
        of Lisbon, Lisboa 1049-001, Portugal \\
        $^5$SKA SA, 3rd Floor, The Park, Park Road, Pinelands, 7405,
        South Africa \\
        $^6$Department of Physics \& Astronomy, University of Pennsylva-
        nia, 209 South 33rd Street, Philadelphia, PA 19104, USA \\
        $^7$NASA Jet Propulsion Laboratory, California Institute of
        Technology, 4800 Oak Grove Drive, MS 169-215, Pasadena, CA,
        91109, U.S.A. \\
        $^8$California Institute of Technology, MC 249-17, Pasadena,
        California, 91125 U.S.A. \\
        $^9$Department of Physics, Blackett Laboratory, Imperial
        College, London SW7 2AZ, UK \\
        E-mail: \email{tcchang@asiaa.sinica.edu.tw}}

\abstract{

We present the science enabled by cross-correlations
of the SKA1-LOW 21-cm EoR surveys with other line mapping programs. In particular, we identify and investigate potential
synergies with planned programs, such as the line intensity mapping of
redshifted CO rotational lines, [CII] and Ly-$\alpha$ emissions during
reionization.  We briefly describe how these tracers of the
star-formation rate at $z \sim 8$ can be modeled jointly before forecasting their
auto- and cross-power spectra measurements with the nominal 21cm EoR
survey. The use of multiple line tracers would be invaluable to validate and enrich our understanding of the EoR.

}

\FullConference{
Advancing Astrophysics with the Square Kilometre Array\\
June 8-13, 2014\\
Giardini Naxos, Italy}

\begin{document}

\section{Introduction}

While the distribution of neutral hydrogen mapped by SKA1-LOW provides an excellent and unique view of the reionization
process over a large range of redshifts, detecting the sources responsible for
reionization directly sheds light on the crucial stage of galaxy
formation and complements our understanding of EoR.  Extremely deep
imaging with the Hubble Space Telescope (HST) has begun to probe the very bright end of the UV
luminosity functions at z > 6 \citep{2014arXiv1403.4295B, 2013ApJ...768...71R}, with improvements expected 
with the  James Webb Space Telescope (JWST). In the sub-mm, the
Atacama Large Millimeter Array (ALMA) has
detected individual high redshift, luminous objects known from
existing surveys (e.g., \cite{2013ApJ...778..102O}). However, observations that are aimed at
detecting individual galaxies at z > 6 are difficult and time
consuming, and neither of these space-borne facilities nor ALMA is
expected to resolve the bulk of low luminosity sources responsible for
reionization \citep{2011MNRAS.414..847S}.  Approaches which can access the entire
luminosity function of reionizing sources are needed.

Line Intensity Mapping has emerged as a promising technique that is
sensitive to the integrated light produced by faint galaxies: instead
of resolving individual sources, one measures on larger spatial scales
the collective emission from an ensemble of sources, while retaining
the spectral--thus redshift--information.  This allows efficient
redshift surveys that probe the integrated luminosity function of
sources and provide three-dimensional information to study star
formation activities during EoR.  

A few spectral lines are currently being
considered as promising tracers for high-redshift star formation
activities in the intensity mapping regime.  Among them, the most
promising ones are the rotational  transitions from carbon monoxide
(CO) \citep{2008A&A...489..489R, 2010JCAP...11..016V, 2011ApJ...730L..30C, 2011ApJ...728L..46G, 2011ApJ...741...70L,
  2013ApJ...768...15P, 2014MNRAS.443.3506B}, the 158$\mu m$ emission from singly ionized
carbon ([CII]) \citep{2012ApJ...745...49G,2014ApJ...793..116U,2014arXiv1410.4808S}, and the Lyman-$\alpha$
transition line from hydrogen \citep{2013ApJ...763..132S,
  2014ApJ...786..111P}.   Such large-scale surveys will not
only reveal early star formation history but also measure the
clustering of ionizing sources.  These line intensity maps mark the
three-dimensional distribution of ionized regions and probe different
gas phases that complements the 21cm EoR surveys which trace neutral hydrogen.
Together, they draw a complete view of the reionization process
in the high-redshift Universe.  In
addition, on scales larger than the ionized regions, these line tracers 
anti-correlate with the 21cm emission.  By measuring the shape of the
cross-power spectrum of the two surveys, one can determine the
characteristic scale
of ionized regions, by marking the scale where the cross-correlation
becomes negative,  and be able to constrain statistically the
characteristic size scale of ionized regions as a function of redshifts \citep{2008A&A...489..489R, 2009ApJ...690..252L}.

Below we discuss each of these tracers in detail, and present
forecasts on the measurements of power spectra and cross-correlation signals with the
SKA1-LOW 21cm EoR survey.  Due to the short emission wavelengths, it is not
possible to observe [CII] and Ly$\alpha$ with the SKA, thus we present
the results assuming other future surveys.  For CO, however, the
proposed highest-frequency band of SKA1-MID can potentially cover the
redshifted CO(1-0) transition at $z>7.3$.  We discuss such
possibilities in the next sections.

\section{CO Intensity Mapping}


The CO(1-0) rotational line has a rest frequency of 115 GHz.  The proposed highest
frequency band of SKA1-MID, 4.6-13.8GHz, will have a chance of
capturing the redshifted CO(1-0) at $z >7.3$.  For intensity
mapping purposes, where we aim to measure the large-scale distribution
of redshifted CO at low-angular resolution, only the inner few kilometer
core of SKA1-MID will be relevant.  The uncertainty
in the theoretical modelling of CO brightness temperature at high
redshifts is still large,
but most predict the amplitude to be about $10^{-6}$K or smaller on
quasi-linear scales.   Large
collecting area and high sensitivity are required,  and a densely packed antenna configuration
at the central core is desired, much like the requirement for SKA1-LOW
for the 21cm EoR survey.  Currently, however, the designed 
filling factor of SKA1-MID at the core is relatively small, only about
$10^{-5}$ in the inner 1 km core in diameter, making the prospect of
detecting redshifted CO very challenging.   

Here we instead use SKA1-MID as a collection of single dishes, where
each antenna records the total power while the cross products between
antenna pairs are ignored.  We note that performing the CO survey in
the single dish mode lifts the compact antenna configuration
requirement and enables a survey in the intensity mapping regime, but
imposes constraints on the stability of the auto-correlation spectrum
which would not be present if it were done interferometrically.  We
consider a narrow redshift range at $z=8\pm0.5$, the lowest possible
redshift allowed by SKA1-MID which should be more CO-rich. The
specification is listed in Table~1, which also specifies the assumed
SKA1-LOW 21cm survey parameters.  Note that the assumed survey areas
for CO auto-power spectrum and CO-HI cross-power spectrum calculations
are chosen to be 0.1 and 10 deg$^2$, respectively, in order to
optimize the signal-to-noise ratio in each case.

\begin{figure}[th]
\begin{center}
\includegraphics*[height=7.2cm]{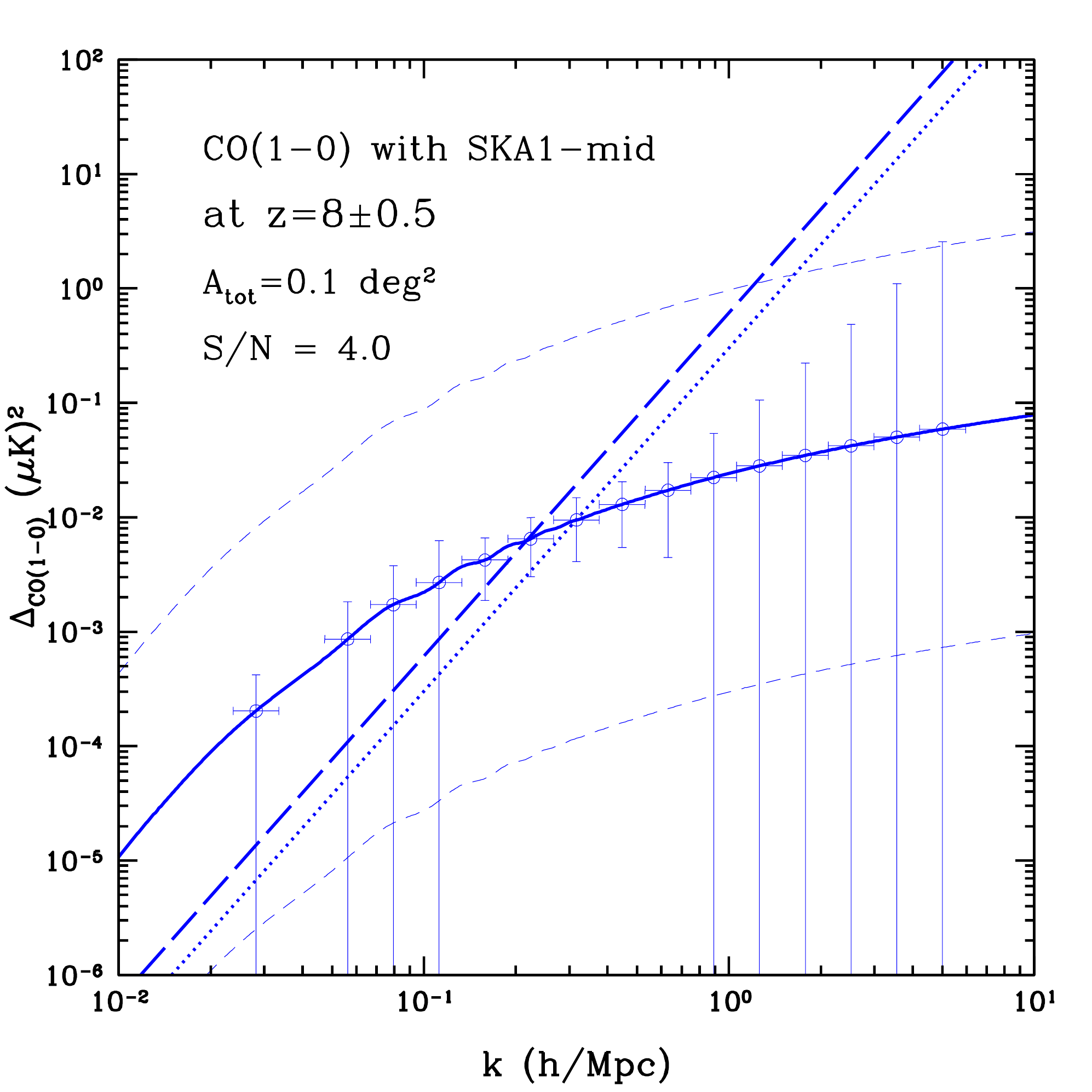}
\includegraphics*[bb=0 0 550 550,height=6.9cm]{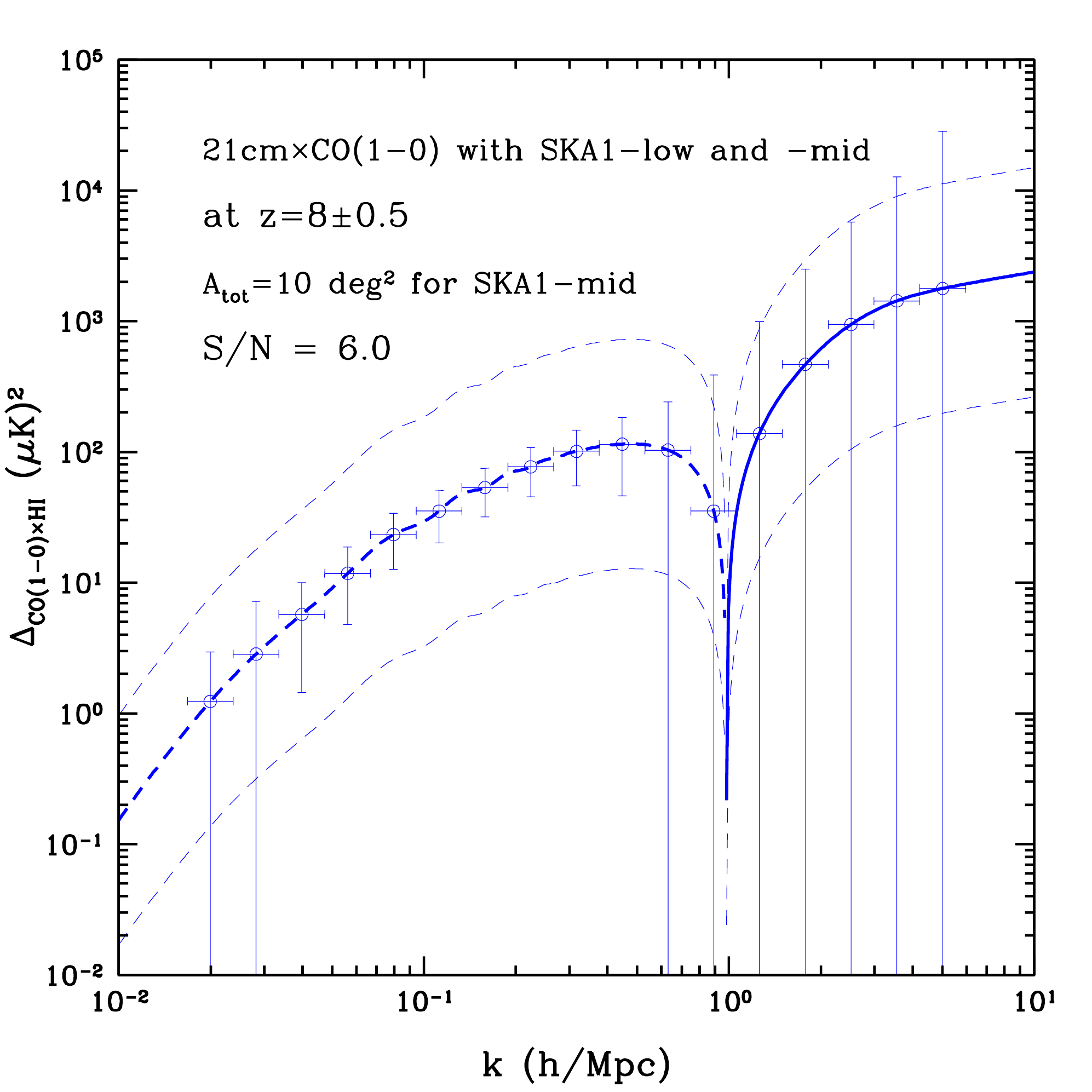}\\
\caption{\textit{Left:} SKA1-MID CO power spectrum at
  $z=8$. \textit{Right:}  SKA1-LOW 21cm x SKA1-MID CO at $z=8$. The
  upper, central, and lower curves indicate predicated CO signal
  strengths from three different models (see texts), whose
  uncertainties are larger than the predicted error bars. The survey areas for CO auto-power spectrum and CO-HI
  cross-power spectrum calculations are assumed to be 0.1 and 10
  deg$^2$, respectively, to optimize the signal-to-noise ratios.}
\label{fig:21xco}
\end{center}
\end{figure}

The amplitude of predicted CO brightness temperature fluctuations is
rather uncertain, differing by orders of magnitude at $z=8$ from model
to model.  This is currently one of the most challenging aspects for
CO intensity mapping work.  Figure~\ref{fig:21xco} shows the
forecasted CO power spectrum and the COx21cm cross-power spectrum
measurements with the survey parameters listed in Table~1.  Here we
demonstrate the model uncertainties by plotting three of them: the
middle solid curve with error bars is based on the models in
\cite{2008A&A...489..489R}, while the upper dashed curve is based on
\cite{2011ApJ...741...70L} and lower dashed curve from
\cite{2011ApJ...728L..46G}. The error bars include contributions from
thermal noise and cosmic variance.  The predicted total
signal-to-noise ratio (SNR), summed over all accessible scales, of the
upper, central and lower power spectra with SKA1-MID in single dish
mode are 20.7, 4.0 and 0.1; for the CO and 21cm cross-power spectra
with SKA1-MID and SKA1-LOW, the corresponding SNRs are 26.1, 6.0 and
0.7.  The CO(1-0) signals and the cross power spectrum between 21cm
and CO(1-0) emissions would be detectable at statistically significant
level in more optimistic scenarios.  The CO model uncertainties,
however, exceed the measurement errors, thus making it difficult to
plan a survey.  On the other hand, a detection of CO power spectrum
can be very discriminating against models and guide our theoretical
understandings.  Observational efforts for mapping CO at modest
redshifts ($z=2-3$) with existing instruments have only reached
initial results, probing the high-end of CO luminosity functions
\citep{2014ApJ...782...78D,2014ApJ...782...79W,2014AAS...22324634H}.
It is essential to improve our theoretical modelling and advance
observational measurements of CO brightness temperature fluctuations
across redshifts, which is currently an active area of research.
In addition, one may need to worry about foreground
contaminations from synchrotron and free-free radiations coming from
the Galaxy and other extragalactic sources. The severity depends on
the observing frequencies (or the redshifted CO rotational lines of interest) and the
CO signal strengths.  However, these foregrounds are expected
to be spectrally smooth and can be separated from the line emissions \citep{2009A&A...501..801A,2014AAS...22313303K}.
At a few tens of gigahertz frequency
range, spinning dust may also be a potential contamination, although
its spectrum and strength are not well known \citep{2013AdAst2013E...2A}.
It will likely require a
dedicated experiment with both a larger field-of-view and greater
surface brightness sensitivity to make a strong CO detection, or a significantly more compact
SKA-MID core than currently envisaged.

\begin{table}[t]
\begin{center}
\caption{Parameters for SKA1-LOW and -MID at z=8$\pm$0.5. For
  SKA1-MID, single dish observation mode using 254 antennae is
  assumed. The survey areas for CO auto-power spectrum and CO-HI
  cross-power spectrum calculations are assumed to be 0.1 and 10
  deg$^2$, respectively, in order to optimize the signal-to-noise ratios. }
\vspace{4mm}
\label{tab:lowpec}
\begin{tabular}{l | c | c | c}
\hline\hline
& SKA1-LOW & SKA1-MID  & unit \\
&& auto/cross  & \\
\hline
ant. diameter $D_{\rm ant}$  & 35 & 15 & m\\
survey area $A_{\rm s}$ & 13 & 0.1/10 & $\rm deg^2$\\
FoV per ant. & 13 & 0.01 & $\rm deg^2$\\
effective area per ant. $A_e$ & 925 & 170 & $\rm m^2$\\
freq. resolution $d\nu$ & 3.9 & 9.7 & kHz\\
bandwidth (z=8$\pm$0.5) BW & 18 & 1427 & MHz \\
tot. int. time $t_{\rm int}$ & 1000 & 10,000 & hr\\ 
min. baseline $D_{\rm min}$ & 30 & - & m\\ 
max. baseline $D_{\rm max}$ & 1 & - & km\\ 
$\rm uv_{\rm min}$ & $16$ & - & \\
$\rm uv_{\rm max}$ & $526$ & - & \\
$\rm T_{\rm sys}$ & 400 & 25 & K\\ 
effective num. ant. & 433 & 254 & \\
num. density of baselines $n_{\rm base}$ & 0.8 & - & \\ 
\hline
\end{tabular}
\end{center}
\end{table}


\section{CII Intensity Mapping}

Carbon is one of the most abundant elements in the Universe and it
becomes singly ionized [CII] with an ionization energy of 11.26 eV, less than that of hydrogen. With a splitting of the fine-structure level at $91\ \rm K$,
[CII] is easily excited resulting
in a line emission at $157.7\ \rm \mu m$ through the $^2P_{3/2}\to
^2P_{1/2}$ transition. It is well established that the bulk of [CII] emission
comes from photodissociation regions (PDRs), and provides a major cooling mechanism
for the neutral interstellar medium (ISM). It is generally the  brightest emission line in star-forming galaxy
spectra and contributes to about 0.1\% to 1\% of the total far-infrared (FIR) luminosity in low redshift galaxies.
Since carbon is naturally produced in stars, [CII] emission is expected
to be a good tracer of the gas distribution in galaxies.  ALMA
high-resolution observations have revealed [CII] in high-redshift galaxies, e.g, \citep{2014arXiv1404.7159R, 2014A&A...565A..59D}, although
no detections have been made for [CII] associated with galaxies at
$z>\sim7$ \citep{2014ApJ...784...99G,2014arXiv1405.5387O} 

Even if the angular 
resolution to resolve the [CII] emission from individual galaxies at high redshift is not available, 
the brightness variations of the [CII] line intensity can be used to map the 
underlying distribution of galaxies and dark matter \citep{2004A&A...416..447B,2010JCAP...11..016V,2012ApJ...745...49G}.

Here we follow \cite{2012ApJ...745...49G} to calculate the expected
[CII] line fluctuations at $ 6 < z < 9$, assuming that [CII] emission
mainly originates in the hot gas in galaxies and that it is
proportional to the gas metallicity, based on both analytical
arguments and numerical models.  Alternatively, [CII] emission can be
estimated using observational relations, such as the empirical relation
between [CII] luminosity and star formation rate (SFR) in low redshift
galaxies from \cite{2012ApJ...755..171S}. This has the advantage of
including [CII] emission from several media, since [CII] is emitted not
only from hot ionized gas but also from cold, mostly neutral gas in PDRs, and also in minor proportion from other
regions. The connection between [CII] emission and SFR can be easily
understood both in PDRs, where the emission of radiation is
proportional to the strength of the far UV radiation field thus to
the SFR, and in ionized regions since their size is proportional to
the ionization rate and thus to the SFR.

\begin{figure}[htbp]
\begin{center}
\includegraphics*[height=7.2cm]{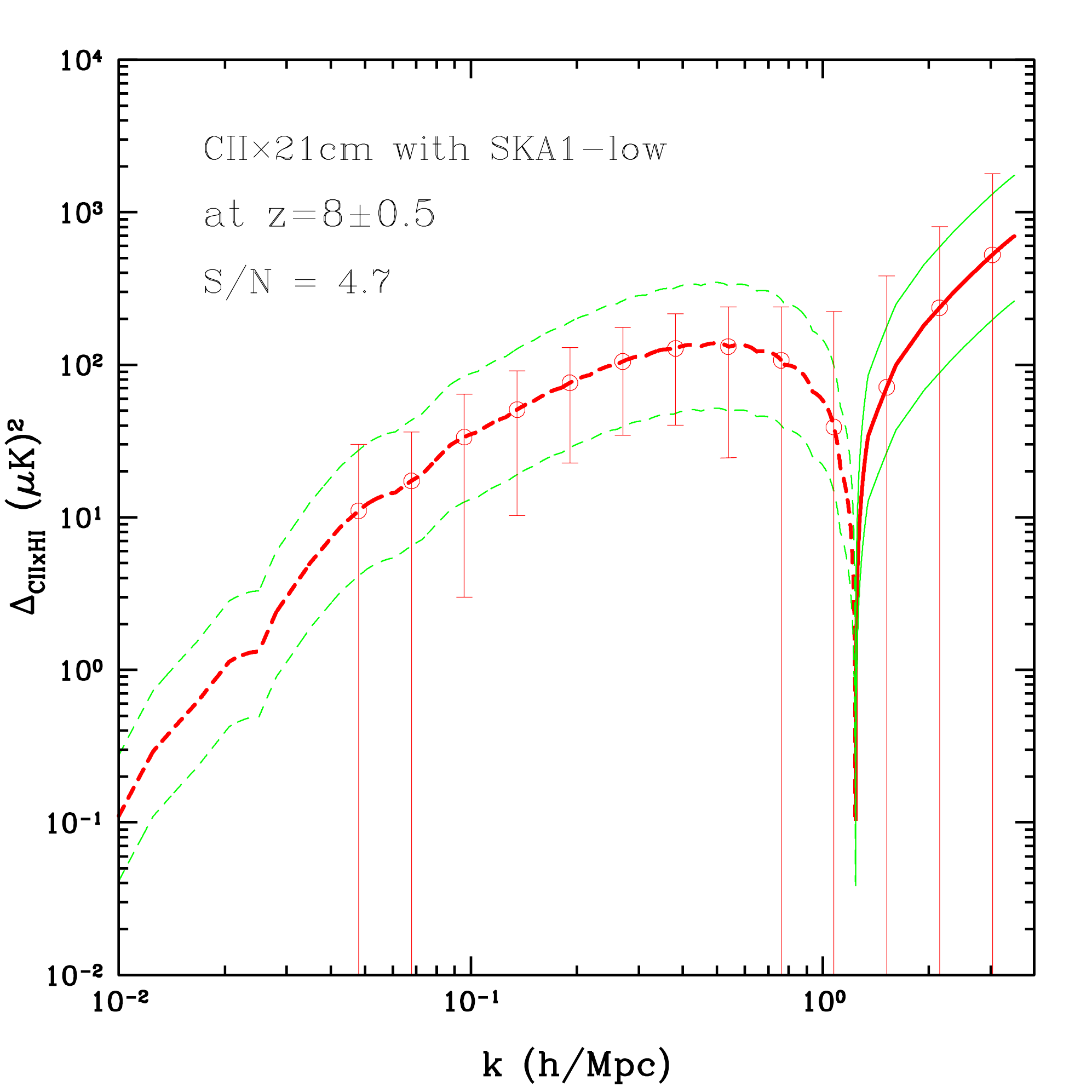}
\caption{The cross-power spectrum of SKA1-LOW 21cm with a potential
  [CII] line mapping program at
  $z=8$. The expected signals are plotted in red, while the theoretical
  uncertainties of the models are indicated by the green curves.  The
  error bars are calculated based on parameters listed in Tables~1 and 2.}
\label{fig:21xcii}
\end{center}
\end{figure}

For redshifted [CII] line, the redshift range
corresponds to observing frequencies of $\sim 200-300$ GHz.  We
assume a future [CII] intensity mapping instrument, based on a grating
spectrometer and 20,000 bolometer
array detectors for spectral line
measurements. The instrument is assumed to be on a telescope with a 10-m
aperture.  The details specifications are listed in Table~2. The forecasted cross-power spectrum of 21cm and [CII] surveys at
$z=8$ is shown in Figure~\ref{fig:21xcii}. The red curve is the predicted amplitude of cross-correlation
based on [CII] models in \cite{2012ApJ...745...49G}, while the green
curves indicate the theoretical uncertainties.  Note the
cross-correlation is negative at $k <\sim 1$ (h/Mpc), a signature from
the anti-correlation of [CII] and 21cm on scales larger than the typical
bubble size, as the former traces star formation activities thus the
ionized region, while 21cm traces the neutral part of the
Intergalactic Medium (IGM).  A 4.7-$\sigma$ detection is expected with this particular setup.
At these frequencies, however, contributions from different CO rotational line emissions coming from different
redshifts may confuse the redshifted [CII] emissions.  One may apply
bright-source masking or template fitting  techniques  to extract the
redshifted [CII] signals.  A pilot [CII] intensity mapping experiment, Time-Pilot, is
currently underway to map out the redshifted [CII] emissions from high
redshifts \citep{2014SPIE.9153E..1WC}.
 
\begin{table*}
\centering                         
\caption{Experimental Parameters for a Possible [CII] Mapping Instrument.}     
\begin{tabular}{l | c c c}   
\hline\hline                 
   Aperture diameter (m) & 10\\
\hline 
   Survey Area ($A_{\rm S}$; deg$^2$) & 16\\
   Total integration time (hours) & 4000\\
   Free spectral range ($B_\nu$; GHz) & $185{-}310$ \\
   Freq. resolution ($\delta_\nu$; GHz) & 0.4 \\
   Number of bolometers & 20,000 \\
   Number of spectral channels & 312 \\
   Number of spatial pixels & 64 \\
   Beam size$^{\mathrm{a}}$ ($\theta_{\rm beam}$; FWHM, arcmin) & $0.4$\\
   Beams per survey area$^{\mathrm{a}}$ & $2.6 \times 10^{5}$ \\ 
   $\sigma_{\rm pix}$: Noise per detector sensitivity$^{\mathrm{a}}$ (Jy$\sqrt{\mathrm{s}}$/sr) & $2.5 \times 10^{6}$ \\
   $t^{\rm obs}_{\rm pix}$: Integration time per beam$^{\mathrm{a}}$ (hours) & 1.0 \\
\hline
$z=8$ $V_{\rm pix}$ (Mpc/h)$^3$ & 4.8 \\
   \hline
   $z=8$  $P^{\rm CII}_N$ (Jy/sr)$^2$ (Mpc/h)$^3$ & 4.3$\times10^9$\\
\hline \hline
\multicolumn{4}{c}{$^{\mathrm{a}}$ values computed at $238 \,$GHz, corresponding to [CII] at $z=7$.}                                
\end{tabular}
\label{tab:cii}     
\end{table*}


\section{Ly$\alpha$ Intensity Mapping}

Ly$\alpha$ photons have a rest-frame wavelength of 1216 $\mathring{A}$ and so Ly$\alpha$ emission during the EoR
will redshift to the near-infrared regime today, making it potentially detectable by narrow-band infrared detectors. Ly$\alpha$ 
photons emitted by galaxies are mostly absorbed and reemitted by the neutral hydrogen in the galaxy 
which causes a scatter of the radiation greatly decreasing the Ly$\alpha$ flux detected by direct observations 
of Ly$\alpha$ emitters. Therefore, galaxy surveys are not able to fully measure all of the intrinsic Ly$\alpha$ 
emission. Intensity mapping is however a low resolution technique and so by not attempting to resolve the 
sources of Ly$\alpha$ photons we can in principle detect all of the Ly$\alpha$ radiation emitted both from galaxies and from the IGM.

Recent work by \cite{2013ApJ...763..132S,2014ApJ...786..111P} describes the process of Ly$\alpha$ emission in the
EoR and post-EoR and shows estimates for intensity mapping of the Ly$\alpha$ signal at redshifts $7 < z < 11$. Ly$\alpha$ 
emission from galaxies is mainly sourced by stellar ionizing radiation since 
stars emit photons which ionize neutral hydrogen, which then emits Ly$\alpha$ photons upon
recombination, and also because the heating of the gas by stellar UV radiation gives rise to e-HI 
collisions causing further Ly$\alpha$ emission. During the EoR stellar populations can also source emission in the dense 
and ionized IGM surrounding the galaxies through e-p recombinations
and e-HI collisions. In addition, diffuse Ly$\alpha$ emission is also
important. It can originate either from e-p recombinations sourced by
X-ray radiation or from stellar continuum photon redshifted into the Ly$\alpha$ line of the IGM. 

The intensity of the contributions from galaxies and from the IGM is dependent on several key astrophysical parameters such as:
the star formation efficiency, the stellar spectrum, the escape fraction of ionizing photons from galaxies to the IGM, the gas
temperature and clumping, the minimum mass of Ly$\alpha$ halos and the ionization state of the IGM. The relative contribution 
from galaxies and from the IGM to the Ly$\alpha$ photons budget is highly dependent on the ionization history 
and on the local heating of the IGM and so it is very difficult to estimate.  

Observational maps of Ly$\alpha$ emission will be contaminated by extragalactic continuum emission and line foregrounds and 
also by emission from our galaxy. Continuum contamination can in principle be removed from intensity maps taking into account 
the smooth evolution of this radiation with frequency compared to the evolution of the Ly$\alpha$ line; however, zodiacal 
light emitted from our galaxy will bring confusion to the observational maps making it possible to only extract the 
Ly$\alpha$ power spectra at small scales where zodiacal light is spatially smooth. Foreground lines from lower 
redshifts, namely the 6563 $\mathring{A}$ H$\alpha$, the 5007 $\mathring{A}$ [OIII] and the 3727 
$\mathring{A}$ [OII] lines will strongly affect Ly$\alpha$ observational maps however their contamination can be removed by 
masking the contaminated observational pixels as was shown in \cite{2014ApJ...785...72G}.

\begin{figure}[htbp]
\begin{center}
\includegraphics*[height=7.2cm]{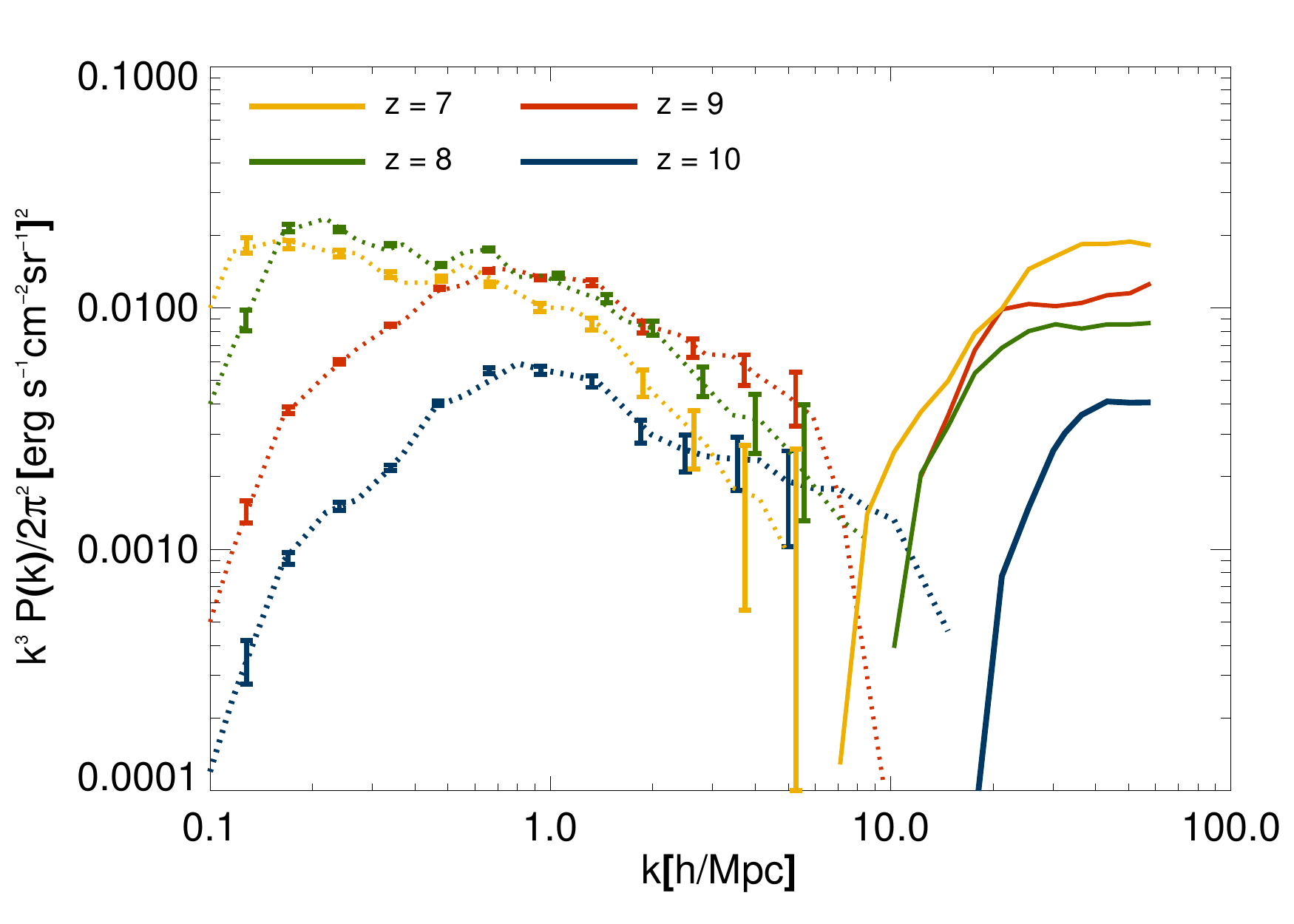}
\caption{The SKA1-LOW 21cm and Ly$\alpha$ cross-power spectra at
  $z=(7, 8, 9, 10)$.  The Ly$\alpha$ models are based on
  \cite{2013ApJ...763..132S}.  The cross-power spectra are predicted to be
  detectable at $>100$ SNR significance level, given the assumed
  survey parameters}.
\label{fig:21xlya}
\end{center}
\end{figure}

Below we illustrate the cross-correlation signature of Ly$\alpha$ and
21cm emissions during EoR, from an assumed Ly$\alpha$ intensity mapping experiment which consists of an aperture 
array with the parameters described in table 3.  The parameters of the
21cm intensity mapping observation with SKA1-LOW are described in Table 1 for $z=8$, and the assumed frequency dependence
of instrument system temperature 
$T_{sys}$ and the collecting area $A_e$ are as follows:  $T_{sys} = T_{sky}
+ T_{rec}$, where $T_{sky} = 60 \left(300 MHz \over \nu \right)^{2.55}
K$ is the sky temperature in Kelvin, and $T_{rec}= 0.1*T_{sky} + 40 K$
is the instrument receiver temperature. $A_e = 925
\left(110 \over \nu \right)^2 m^2$.   The Ly$\alpha$ intensity mapping
calculation is from \cite{2013ApJ...763..132S}. The forecast shows
that the 21cm and Ly$\alpha$ cross-power spectra can be detected at high
SNR values of (789, 442, 273, 462) for $z=(7,8,9,10)$, respectively,
and is shown in Figure~\ref{fig:21xlya}.

\begin{table*}
\centering                         
\caption{Experimental Parameters for a Possible Ly$\alpha$ Mapping Instrument.}     
\begin{tabular}{l | c c c}   
\hline\hline                 
   Aperture diameter (m) & $0.2$\\
\hline 
   Survey Area ($A_{\rm S}$; deg$^2$) & $13$\\
   Total integration time (hours) & $2900$\\
   Free spectral range ($B_{\lambda}$; $\mu$ m) & $0.85{-}1.1$ \\
   Freq. resolution ($\lambda/\delta_{\lambda}$) & $220$ \\
   Number of pixels in 2D array &  $72900$\\
   FOV per pointing; deg$^2$  & $0.6$ \\ 
   Observational time per pointing (hours) & $129.5$ \\
   Survey volume (Mpc/h)$^3$ & $8.5\times10^7$\\
\hline \hline                
\end{tabular}
\label{tab:cii}     
\end{table*}



\section{Discussion}

The strength of the correlation between HI and molecular line emission
depends on multiple factors and in particular, the sign of the
correlation will depend on whether the emission comes mostly from the
galaxies or the IGM. When cross-correlating CO with HI, the
cross-correlation signals are expected to be associated with the clustering of galaxies (CO) and the IGM
(HI), thus there is a strong and easy to interpret anti-correlation. When cross-correlating
[CII] with HI, the bulk of the [CII] emissions are expected to come
from galaxies, with some small contributions from the IGM, and we
expect a strong anti-correlation. At the high redshifts of interest
($z$>6), a small amount of metals may reside in the IGM, however, the [CII]
spin temperature is expected to follow the CMB temperature
\citep{2012ApJ...745...49G} thus little [CII] emission from the IGM is expected.  For 
cross-correlations of HI with Ly$\alpha$ emissions, the situation is more
complicated since the Ly$\alpha$ emission from the IGM can be very high and
is very uncertain.  Depending on the model considered, the
emission from the IGM can even be higher than the emission from
galaxies at some redshifts. In this case, Ly$\alpha$ and HI would be
positively correlated on large scales; on the other hand, if no
Ly$\alpha$-HI cross-correlation signals are found, we
can place constraints on the Ly$\alpha$ sources of emission. 

\section{Summary}

In this chapter we motivated a novel use of SKA1-LOW 21cm to probe
the epoch of reionization. By cross-correlating the 21cm line with
other atomic and molecular lines observed in the intensity mapping regime, we can
not only validate a potential 21cm EoR detection but also learn more about the EoR
itself. On the one hand, the 21cm line traces neutral regions, the
yet to be ionized universe while the other atomic lines trace star
formation activities. The combination thus forms a potent, complete and
unique picture of the reionization process.

We focus our study on CO, [CII] and Ly$\alpha$ transition lines currently
identified as the most promising tracers. For CO, we
study the possibility of using directly SKA1-MID at appropriate frequencies to map the EoR at z>7.5. For [CII] and Ly$\alpha$, we assume
the successful deployment of relevant instruments currently being planned. In
all cases, we find that the strength of our detection strongly varies
with theoretical model. For example, in the case of CO, according to
the particular models considered, we could go from a strong detection (greater than 6
$\sigma$) to a non-detection. The same holds for [CII], and it appears
more promising for Ly$\alpha$ to be detected at high SNRs . While this situation might be worrisome, it simply reflects
the fact that we are probing a totally new territory for
astrophysics.  Besides, these line tracers have different sources of astrophysical
contamination which are in general much less severe than  the ones
plaguing the redshifted 21 cm line. 
A cross-correlation measurement can thus serve as an independent confirmation of the cosmological
origin of the measured signals.  The use of multiple line tracers would thus be invaluable to
validate and enrich our understanding of the EoR, and it will open up a
huge discovery space.


\bibliographystyle{apj}
\bibliography{synergy.bib}

\end{document}